\documentclass[conference]{IEEEtran}
\IEEEoverridecommandlockouts
\usepackage[nospace,noadjust]{cite}
\usepackage[dvipsnames]{xcolor}
\usepackage{amsmath,amssymb,amsfonts}
\usepackage{amsthm}
\usepackage{graphicx,caption}
\usepackage{bbm}
\usepackage{setspace}
\usepackage{braket}
\usepackage{algpseudocode}
\usepackage[]{algorithm}
\usepackage{textcomp}
\usepackage{xcolor}
\usepackage{soul}
\usepackage{caption}
\def\BibTeX{{\rm B\kern-.05em{\sc i\kern-.025em b}\kern-.08em
		T\kern-.1667em\lower.7ex\hbox{E}\kern-.125emX}}
\usepackage{nccmath}
\usepackage{array}
\usepackage{subfig}
\usepackage{bigstrut}
\usepackage{booktabs}
\definecolor{wblue}{RGB}{0,58,199}
\usepackage{setspace}

\usepackage{multicol}
\usepackage[T1]{fontenc}
\usepackage[utf8]{inputenc}
\usepackage{babel}
\usepackage[font=small,labelfont=bf]{caption}
\usepackage{bm}
\usepackage{blindtext}
\usepackage{caption}
\usepackage{bigstrut}
\newtheorem{thm}{Theorem}
\newtheorem{lem}[thm]{Lemma}

\algnewcommand{\IIf}[1]{\State\algorithmicif\ #1\ \algorithmicthen}
\algnewcommand{\EndIIf}{\unskip\ \algorithmicend\ \algorithmicif}
\algnewcommand{\FFor}[1]{\State\algorithmicfor\ #1\ \algorithmicdo}
\algnewcommand{\EndFFor}{\unskip\ \algorithmicend\ \algorithmicfor}

\ifCLASSINFOpdf
\else
\fi

\begin{document}
	%
	\title{
		{{QoS-Aware Sum Capacity Maximization for Mobile Internet of Things Devices Served by UAVs}}
		{\footnotesize \textsuperscript{}}}
	
	\author{\IEEEauthorblockN{Mohammadsaleh Nikooroo\textsuperscript{1}, Zdenek Becvar\textsuperscript{1}, Omid Esrafilian\textsuperscript{2},  David Gesbert\textsuperscript{2}}
		\IEEEauthorblockA{\textit{\textsuperscript{1} Faculty of Electrical Engineering} 
			\textit{Czech Technical University in Prague},	Prague, Czech Republic \\
			\textsuperscript{2}	\textit{Communication Systems Department, EURECOM}, Sophia Antipolis, France\\
			\textsuperscript{1}\{nikoomoh,zdenek.becvar\}@fel.cvut.cz, \textsuperscript{2}\{esrafili,gesbert\}@eurecom.fr }


		%
		%
		
		\thanks{\textcolor{black}{This work was supported by the project No. LTT 20004 funded by Ministry of Education, Youth and Sports, Czech Republic and by the grant of Czech Technical University in Prague No. SGS20/169/OHK3/3T/13, and partially by the HUAWEI France supported Chair on Future Wireless Networks at EURECOM.}}}

	\maketitle
	
	\begin{abstract}
		The use of unmanned aerial vehicles (UAVs) acting as flying base stations (FlyBSs) is considered as an effective tool to improve performance of the mobile networks. Nevertheless, such potential improvement requires an efficient positioning of the FlyBS. In this paper, we maximize the sum downlink capacity of the mobile Internet of Things devices (IoTD) served by the FlyBSs while a minimum required capacity to every device is guaranteed. To this end, we propose a geometrical approach allowing to derive the 3D positions of the FlyBS over time as the IoTDs move and we determine the transmission power allocation for the IoTDs. The problem is formulated and solved under practical constraints on the FlyBS’s transmission and propulsion power consumption as well as on flying speed. The proposed solution is of a low complexity and increases the sum capacity by 15\%-46\% comparing to state-of-the-art works.  
	\end{abstract}

	
	\section{Introduction}\label{sec:1}
	\par
	Deployment of unmanned aerial vehicles (UAVs) acting as flying base stations (FlyBSs) is a promising way to improve performance in 6G mobile networks, since the FlyBSs offer a high mobility and an adaptability to the environment via flexible movement in 3D. Potential benefits offered by the FlyBSs, however, comes along with challenges related to radio resource management and positioning of the FlyBSs \cite{Li2019}, \cite{Mach2022}, \cite{Nikooroo2022TNSE}, \cite{Mozaffari2016}, \cite{Spyridis2021}, \cite{Esrafilian}.

	The problem of the FlyBS’s positioning is investigated in many recent works. The objectives targeted in those works include a maximization of the downlink sum capacity  \cite{Ahmed2020}, a maximization of the minimum capacity \cite{Ji2020},  a maximization of the uplink capacity \cite{Wei2020}, a maximization of the sum of uplink and downlink capacities \cite{Hua2020_TCOM}, \cite{Hua2020_WCL}, a maximization of the minimum average capacity for device-to-device communication \cite{Li2021}, a maximization of the minimum capacity in networks of sensors or Internet of Things devices (IoTDs) \cite{Xie2020}, a minimization of the FlyBS's power consumption \cite{Tun2021}, a minimization of the number of FlyBSs to guarantee users' QoS requirements \cite{Shi2020}. However, the users considered in \textcolor{black}{\cite{Ahmed2020}-\cite{Shi2020}} are static (i.e., do not change their location over time). \textcolor{black}{This is a required assumption in the solutions provided by those works in \cite{Ahmed2020}-\cite{Shi2020}, and the FlyBS's entire trajectory is derived before the beginning of mission knowing that the users do not move during the mission. An extension of the solutions in these papers to the scenario with moving users is not straightforward.} Furthermore, a guarantee of the minimum capacity to the users is not considered in \cite{Ahmed2020}, \cite{Hua2020_TCOM},  \cite{Hua2020_WCL}, hence, the solutions cannot be adopted in applications, where the quality of service is concerned.

	A solution potentially applicable to the scenarios with moving users is outlined in \cite{Ishigami2020}, where the FlyBSs’ altitude is optimized to maximize the average  system throughput. Then, in \cite{Chen2020}, the authors optimize the number of FlyBSs and their positions to maximize sum capacity. However, neither \cite{Ishigami2020} nor \cite{Chen2020} provide any guarantee of the minimum capacity to the users. In \cite{Zhang2021}, the sum capacity is maximized via a positioning of the FlyBSs using reinforcement learning. Furthermore, the problem of the transmission power allocation is investigated in \cite{Muntaha2021} to maximize the energy efficiency, i.e., the ratio of the sum  capacity to the total transmission power consumption. The minimum required capacity in \cite{Zhang2021} and \cite{Muntaha2021} is assumed to be equal for all users. Besides, the FlyBS's positioning is not addressed in \cite{Muntaha2021} at all and the transmission power allocation is not considered in \cite{Zhang2021}. 
	Then, the minimum capacity of the users is maximized via the FlyBS’s positioning and the transmission power allocation in \cite{Valiulahi2020}. Nevertheless, no constraint on the FlyBS’s speed is considered. 
	
	
	Surprisingly, there is no work targeting the sum capacity maximization in a practical scenario with \textit{moving sensors/IoTDs} and with the \textit{minimum capacity guaranteed to the individual sensors/IoTDs}. 
	All related works are either focused on the scenario where data is collected from static users with apriori known coordinates or no minimum capacity is guaranteed to the users. 
	We target the scenario with mobile devices and the minimum capacity guarantee and we propose a low-complexity solution based on an alternating optimization of the FlyBS’s positioning and the transmission power allocation to the devices. The proposed optimization is done with respect to the feasibility region that is derived via a proposed geometrical approach. With respect to a majority of the related works, we also consider practical aspects and constraints of the FlyBSs including limits on the flying speed, transmission power, and propulsion power.

	The rest of this paper is organized as follows. In Section II we provide the system model for FlyBS-enabled sensor network and we formulate the problem of sum capacity maximization. Next, we propose a method to check the feasibility of a solution to the FlyBS’s positioning and transmission power allocation in Section III. 
	Then, we propose a solution based on an alternating optimization of the transmission power allocation and the FlyBS's positioning in section IV. A geometrical approach is proposed for the FlyBS's positioning. Then, in section V, the adopted simulation scenario and parameters are specified and the performance of our proposed solution is shown and compared with state-of-the art schemes. Last, we conclude the paper and outline the potential future extensions in Section VI.
	
	
	\section{System model and problem formulation}\label{sec:2}
	
	In this section, we first define the system model. Then, we formulate the constrained problem of the sum capacity maximization.
	
\begin{figure}[!t]
		\centering
		\includegraphics[width=3.3in]{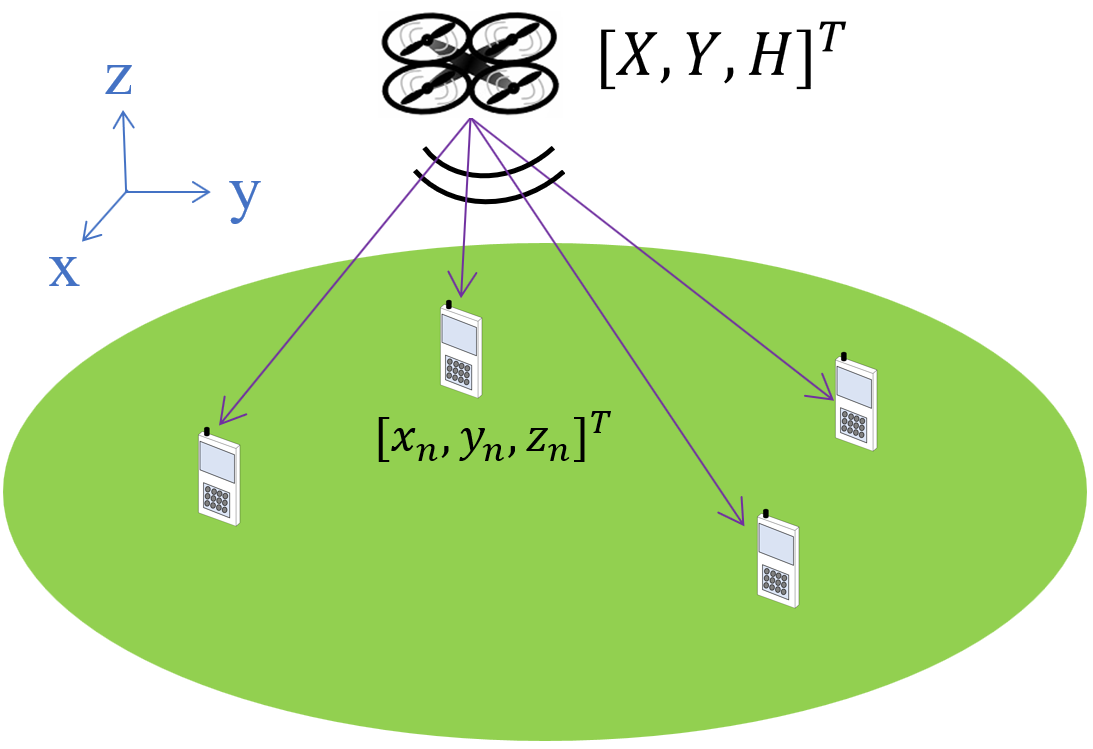}	
		\captionsetup{justification=centering}
		\caption{System model with mobile IoTDs placed within the coverage area of the FlyBS.} 
		\label{fig:sysmodel}
	\end{figure}

	In our system model, one FlyBS serves $ N $ sensors/IoTDs $ \{u_1,...,u_N\} $ in an area as shown in Fig. \ref{fig:sysmodel}. Let $ \bm{q}(t)=\big[X[k], Y[k], H[k]\big]^T $ denote the location of the FlyBS at time step $ k $. We refer to the IoTDs/sensors as nodes in the rest of this paper. Let $ \bm{v_i}[k]=\big[x_i[k],\ y_i[k],z_i[k]\big]^T $ denote the coordinates of node $ i $ at time step $ k $. Then, $ d_i[k] $ denotes Euclidian distance of node  $ i $ to the FlyBS at time step $ k $.

	\vspace{0\baselineskip}

	We adopt orthogonal downlink channel allocation for all nodes. Thus, the channel capacity of node $ i $ is:
	
	\begin{gather}
	C_i[k]=B_i\log_2{\left(1+\frac{p_i^R[k]}{N_i+I}\right)},
	\label{eqn:1}
	\end{gather}
	
	\noindent where $ B_i $ denotes the bandwidth of the $ i $-th node’s channel (note that $B_i$ can differ among nodes), $ N_i $ is the noise power at the $i$-th node's channel, $ I $ denotes the background interference from neighboring base stations (both flying and static), and $ p_i^R[k] $ is the received  power by the $ i $-th node at time step $ k $.

	Let  $ \bm{p}^T=[p_1^T,...,p_N^T] $ denote the FlyBS’s transmission power allocated to all $ N $ nodes. According to the Friis’ transmission equation, the received signal's power at node $i$ ($i\in [1,N]$) from the FlyBS is calculated as: 
	
	\begin{gather}
	p_i^R[k]={Q_i(\frac{\gamma}{\gamma+1}\overline{h}_i +\frac{1}{\gamma+1}\tilde{h}_i )}p_i^T[k]{d_i}^{-\alpha_i}[k],
	\label{eqn:2}
	\end{gather}

	\noindent where the coefficient $ Q_{i} $ is a parameter depending on the communication frequency and gain of antennas. Furthermore, $ \gamma $ is the Rician fading factor, $ \overline{h}_i $ is the line-of-sight (LoS) component satisfying $ |h_n |= 1$, and $\tilde{h}_i$ denotes the non-line-of-sight (NLoS) component satisfying $ \tilde{h}_i \sim $ $ CN(0,1) $, and $ \alpha_{i} $ is the pathloss exponent of the channel for node $i$.

	
	
	\textcolor{black}{For the propulsion power consumption, we refer to the model provided in \cite{Zeng2019} for rotary-wing UAVs. More specifically, the propulsion power is expressed as:}
	\textcolor{black}{{\begin{gather}
			P_{pr}[k]=L_{0}\big(1+\frac{3V_{F}^{2}[k]}{U_{\text{tip}}^{2}}\big)+\frac{\eta_0\rho s_rAV_{F}^3[k]}{2}+\nonumber\\L_{i}\big(\sqrt{1+\frac{V_{F}^4[k]}{4v^4_{0,h}}}-\frac{V_{F}^{2}[k]}{2v^2_{0,h}}\big)^{\frac{1}{2}},\label{eqn:PropulsionModel}
			\end{gather}}
	}
	
	\noindent \textcolor{black}{where $V_F[k]$ is the FlyBS's speed at the time step $k$. Furthermore, $ L_0 $ and $ L_i $ are the blade profile and induced powers in hovering status, respectively, $ U_{\text{tip}} $ is the tip speed of the rotor blade, $ v_{0,h} $ is the mean rotor induced velocity during hovering, $ {\eta_0} $ is the fuselage drag ratio, $ \rho $ is the air density, $ s_r $ is the rotor solidity,  and $ A $ is the rotor disc area. }
	
	
	Our goal is to find the position of the FlyBS to maximize the sum capacity at every time step $ k $ while the node’s minimum required capacity is always guaranteed with practical constraints implied by FlyBSs. Hence, we formulate the problem of the sum capacity maximization as follows:
	\begin{gather}
	\operatorname*{max}_{\bm{{p}}^{\bm{T}}[k], \bm{q}[k]} C_{tot}[k],\forall k,\quad\quad\quad\quad\quad\quad\quad\quad\quad\quad\label{eqn:3}\\[-0pt]
	\text{s.t.}\quad C_i[k]\geq C_i^{min}[k], \quad i\in [1,N],  \quad  (\text{4a})\nonumber\\[-0pt]
	H_{min}[k]\leq H[k]\leq H_{max}[k], \quad\quad\quad\quad (\text{4b})\nonumber\\
	||\bm{q}[k]-\bm{q}[k-1]||\le V_{F}^{max}\delta_k,\quad\quad\quad{\;}{\;}  (\text{4c})\nonumber\\[-0pt]
	P_{pr}[k]\leq P_{pr,th}[k],\quad\quad\quad\quad\quad\quad\quad\quad{\;}  (\text{4d})\nonumber\\
	\sum\nolimits_{i=1}^{N}p_i^T[k]\le p_{max}^T, \quad p_i^T[k]\geq 0, 	\quad{\;}{\;} (\text{4e}) \nonumber
	\end{gather}
	
	\noindent where $ C_{tot}[k]=\sum_{i=1}^{N}{C_i[k]} $ is the sum capacity of the nodes at time step $ k $, 
	$ C_i^{min}[k] $ denotes the minimum capacity required by node $ i $ at time step $ k $, $H_{min}$ and $H_{max}$ are the minimum and maximum allowed flying altitude of the FlyBS at the time step $k$, respectively, and are set according to the environment as well as the flying regulations. Furthermore, $ V_{F}^{max} $ is the FlyBS’s maximum supported speed, $\delta_k$ is the duration between the time steps $k-1$ and $k$, $ ||.|| $ is the $  \mathcal{ L}_2 $ norm, and $ p_{max}^T $ is the FlyBS’s maximum transmission power limit. The constraint ($ \text{4a} $) ensures that every node always receives the required capacity. Furthermore, ($ \text{4b} $) and ($ \text{4c}$) restrict the FlyBS's speed to the range of $[0,V_{F}^{max}] $ and $\big[H_{min}[k],H_{max}[k]\big]$, respectively. In addition, the constraints  ($ \text{4d}$) and ($ \text{4e}$) assure that the FlyBS's propulsion power and total transmission power would not exceed $P_{pr,th}$ and $ p_{max}^T $, respectively. In practice, the value of $P_{pr,th}$ can be set/adjusted at every time step and according to available remaining energy in the FlyBS's battery to prolong the FlyBS's operation.

	Challenging aspects to solve (\ref{eqn:3}) include: $ i $) before the positioning of the FlyBS, a feasibility of the solution to (3) should be verified due the constraints (4a)-(4e), and  $ ii $) the objective function $ C_{tot} $ and the constraint (4e) are non-convex with respect to $ \bm{q} $.

	To tackle the aspect $ i $), we propose a geometrical approach with a low complexity to check the feasibility of any solution to (\ref{eqn:3}). If there is a feasible solution, the proposed approach further determines the feasibility domain used for a derivation of the FlyBS’s positions. To tackle the aspect $ ii $), we propose a suboptimal solution using an alternating optimization of the transmission power allocation and the FlyBS’s positioning based on a local approximation of the objective function. In particular, we propose an iterative approach based on two steps: $1)$ an optimization of the transmission power allocation $ \bm{p^T} $ at the given position of the FlyBS and, $2)$ an update (optimization) of the FlyBS’s position for the derived vector $ \bm{p^T} $ from the step $1$ via a consideration of the feasibility domain defined by the constraints in (\ref{eqn:3}).
	We elaborate the derivation of feasibility domain in Section \ref{sec:3}. Then, we explain our proposed alternating optimization of the transmission power and the FlyBS’s positioning in Section \ref{sec:4}. 
	
	
	\section{Feasibility of a solution}\label{sec:3}
	
	In this section, we present a geometrical approach to check the feasibility of an arbitrary solution to (\ref{eqn:3}) via a consideration of the constraints in (\ref{eqn:3}). 
	Let us first rewrite the constraint ($ \text{4a} $) for an arbitrary setting of the transmission power allocation $ \bm{p^T} $ to individual nodes by means of (\ref{eqn:1}) and (\ref{eqn:2}) as follows:
	\begin{gather}
	C_i=B_i\log_2{\left(1+\frac{Q_ip_i^T}{{{d}_i}^{\alpha_i}(N_i+I)}\right)}\geq C_i^{min},
	\label{eqn:4}
	\end{gather}
	
	\noindent which yields
	\begin{gather}
	d_i\le{(\frac{Q_ip_i^T}{(2^\frac{C_i^{min}}{B_i}-1)(N_i+I)})}^\frac{1}{\alpha_i}=\rho_i,\quad  1\le i\le N.
	\label{eqn:5}
	\end{gather}
	
	Each of the $ N $ inequalities in (\ref{eqn:5}) demarcates a sphere in 3D space. In particular, for every $ i\in [1,N] $, the inequality in (\ref{eqn:5}) implies that the FlyBS lies inside or on the sphere with a center at the location of node $ i $ and with a radius of $ \rho_i $. 
	
\begin{figure}[!t]
		\centering
		\includegraphics[width=2.35in]{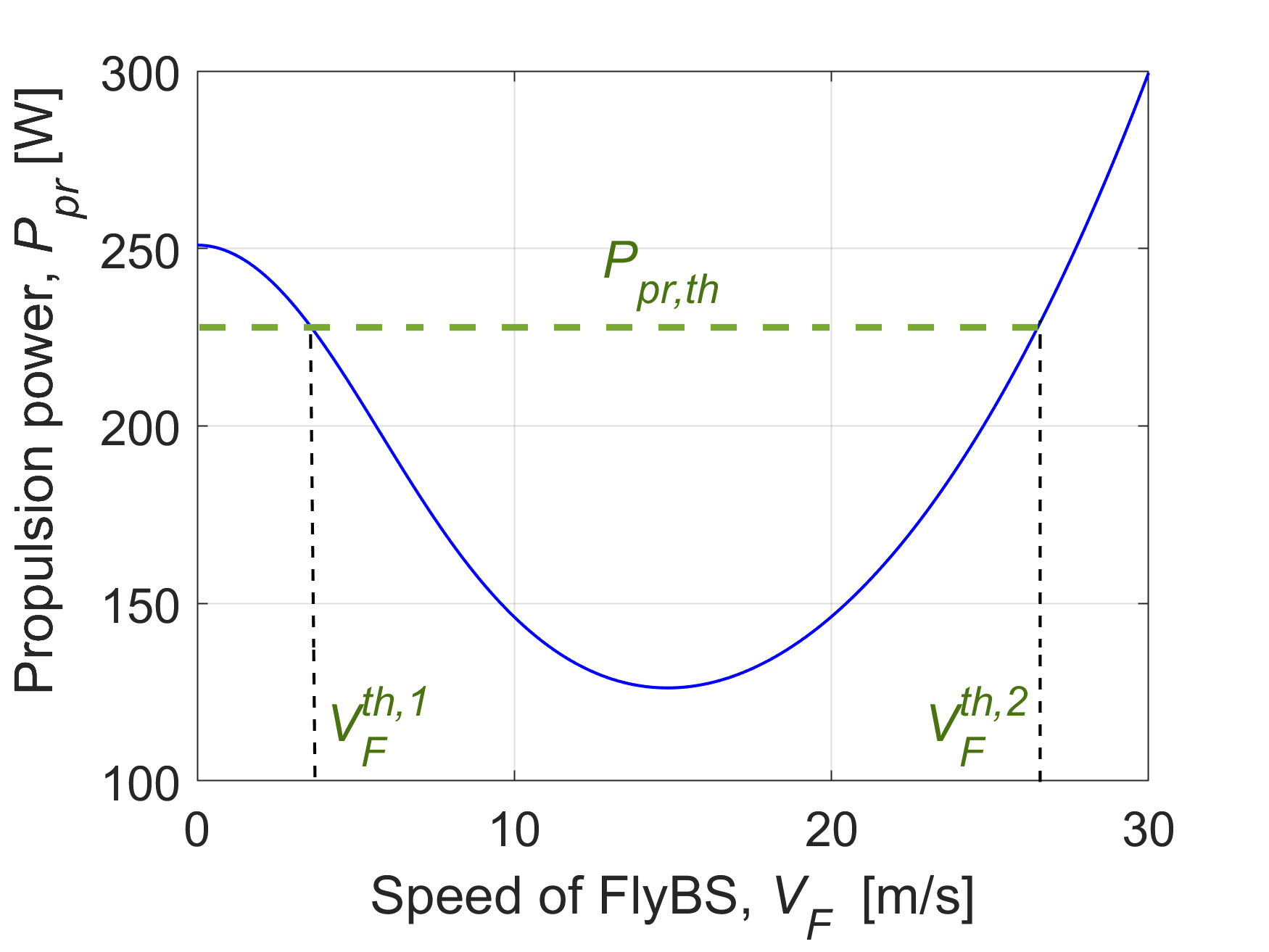}	
		\caption{\textcolor{black}{Propulsion power model vs. speed for rotary-wing FlyBS. }} 
		\label{fig:propulsion_model}
	\end{figure}

	Next, the constraint ($\text{4b}$) defines the next position of the FlyBS on or between the planes $z=H_{min}[k]$ and $z=H_{max}[k]$. In addition, according to Fig. \ref{fig:propulsion_model}, the constraint (4d) is translated as $V_F\in [V_F^{th,1},V_F^{th,2}]$. By combining this inequality with (4c) we get
	\begin{gather}
	||\bm{q}[k]-\bm{q}[k-1]||\le (\text{min}\{V_{F,max},V_F^{th,2}\})\delta_k, \label{eqn:4c-4d_combined_1}
	\end{gather}
	and
	\begin{gather}
	||\bm{q}[k]-\bm{q}[k-1]||\geq V_F^{th,1}\delta_k.\label{eqn:4c-4d_combined_2}
	\end{gather}
	\noindent The equations (\ref{eqn:4c-4d_combined_1}) and (\ref{eqn:4c-4d_combined_2}) define the FlyBS’s next possible position as the border or inside of a region enclosed by two spheres centered at $ \bm{q}[k-1] $ (i.e., the FlyBS’s position at the previous time step) and with radii of  $ V_F^{th,1}\delta_k$ and ($\text{min}\{V_{F,max},V_F^{th,2}\})\delta_k$. Furthermore, to interpret the constraint (4e) in terms of the FlyBS’s position, in following Lemma \ref{lem1}, we derive a necessary condition for the FlyBS’s next position so that there exists a feasible position of the FlyBS for an arbitrary setting of $ \bm{p^T} $.
	
	\vspace{0\baselineskip}

	\begin{lem}\label{lem1}
		For the power allocation vector $ \bm{p^T} $ at time step $ k-1$, a necessary condition for a feasibility of any solution to the positioning of the FlyBS at time step $ k $ is: 
		\begin{gather}
		\big|\big|\bm{q}[k]-{\bm{\theta_0}}({\bm{p}}^{\bm{T}},k)\big|\big|\le\Upsilon({\bm{p}}^{\bm{T}},k),
		\label{eqn:6}
		\end{gather}
		
		\noindent where $ \bm{\theta_0}(\bm{p}^{\bm{T}},k)=[\frac{\sum_{i=1}^{N}\iota_ix_i}{\sum_{i=1}^{N}\iota_i},\frac{\sum_{i=1}^{N}\iota_iy_i}{\sum_{i=1}^{N}\iota_i},H] $, and $\iota_i$ is a substitution derived  in the proof, and $ \Upsilon(\bm{p}^{\bm{T}},k)={(\frac{p_{max}^T-\chi}{\sum_{i=1}^{N}\iota_i})}^\frac{1}{2} $.  
		
	\end{lem}
	
	\textit{proof. Please see Appendix \ref{sec:apxA}}

	Note that, an existence of a feasible solution is contingent upon all the constraints in (\ref{eqn:3}) and not only the condition (\ref{eqn:6}). Thus, we now analyze the feasibility of any solution to (\ref{eqn:3}) by incorporating the constraints derived for the FlyBS’s next position. In order to check if these inequalities hold at the same time, we propose the following low-complexity approach. Let $ {cl}_i$ $(i\in[1,+2]) $ denote $ N+2 $ spheres defined by the inequalities (\ref{eqn:5}), (\ref{eqn:4c-4d_combined_1}), and (\ref{eqn:6}). Note that we deal with (\ref{eqn:4c-4d_combined_1}) later in this section. The $ N $ spheres represented by (\ref{eqn:5}) have centers at the same position as their corresponding nodes. Furthermore, the sphere indicated by  (\ref{eqn:4c-4d_combined_1}) has a center at the FlyBS’s position at time step $ k-1 $.
	
	Then, for each pair of spheres $ {sp}_j $ and $ {sp}_k $, we consider their intersection. There are three different cases regarding the intersection: $ i $) $ {sp}_j  $ and $ {sp}_k $ have no intersection point and lie completely outside each other, $ ii $) $ {sp}_j $ and $ {sp}_k $ have no intersection points and one of these spheres lies inside the other one, $ iii $) $ {sp}_j $ and $ {sp}_k $ intersect on their borders which is in the shape of a circle (assuming that a single point is also a circle with radius of zero). Note that any two spheres from the set of spheres indicated by (4a) do not completely overlap, as each sphere has a distinct center. Furthermore, if any sphere represented by (\ref{eqn:4c-4d_combined_1}) or (\ref{eqn:6}) is identical to another sphere, we simply ignore one of those spheres.  
	
	For the case $ i $, we conclude that at least two of the constraints in (\ref{eqn:3}) do not hold at the same time and, thus, there is no feasible solution to (\ref{eqn:3}). For the case $ ii $, one of the constraints in (\ref{eqn:3}) corresponding to the outer sphere is automatically fulfilled if the other constraint (the one corresponding to the inner sphere) holds. In such case, we ignore the constraint corresponding to the outer sphere and the rest of the constraints are dealt with according to case $i$ or $iii$. For the case $ iii $, we propose the following low-complexity method to verify the non-emptiness of the intersections of the spheres (in other words, a feasibility of a solution to (\ref{eqn:3})): given the fact that the intersection of a plane and a sphere is circle (if not empty), we search for the intersection of the spheres only on certain planes. In particular, corresponding to each of the $N+2$ spheres $ {sp}_j $, consider two horizontal planes ${pl}_{j,1}$ and ${pl}_{j,2}$ that are tangent to $ {sp}_j $ (one at the topmost point on $ {sp}_j $ and one at the lower most point). Then, we remove from the set of the derived planes those that do not fulfill the altitude constraint (4b). Hence, at most $2N+4$ horizontal planes are derived. Next, for each of the remaining planes, we find the intersection of the plane and all the spheres. Let ${cl}_{l,k,1}$ and ${cl}_{j,k,2}$ be the intersection (circle) of $ {sp}_k $ on ${pl}_{j,1}$ and on ${pl}_{j,2}$, respectively. On each plane, we derive and collect the intersection points of each two of such circles. Then, we verify whether there are any points in the set of the collected points that would lie inside or on the border of all the circles on the same plane. In case that there are no such points on any of the planes, there is no feasible solution to (\ref{eqn:3}) as all the constraints in (\ref{eqn:3}) cannot be met at the same time. Otherwise, there would be a solution if the remaining condition (\ref{eqn:4c-4d_combined_2}) is also met for at least one of those eligible candidate points.  
	From the described process, the computational complexity of the proposed feasibility check scales as $ (2N+4)\times \binom{N+2}{2}\times (N+2) $, i.e., it is $ O(N^4) $.  
	
	In the next section, we target the problem of power allocation and the FlyBS’s positioning in (\ref{eqn:3}) and we show how the FlyBS's position is determined with respect to the constraint spheres $ {sp}_j $ ($ j\in [1,N+2] $) derived in this section. 

	\section{\textcolor{green}{} FlyBS positioning and power allocation}\label{sec:4}

	In this section, we outline our proposed FlyBS’s positioning and transmission power allocation maximizing the sum capacity under the feasibility condition derived in Section \ref{sec:3}.

	Our proposed solution is based on alternating optimization updating the transmission power $ \bm{p^T} $ and the FlyBS’s position $ \bm{q} $ at every time step. First, note that for a given $ \bm{q} $, the problem of the $ \bm{p^T} $ optimization is solved via CVX, as the sum capacity in (\ref{eqn:3}) is concave, and the constraints in (\ref{eqn:3}) are convex with respect to $ \bm{p^T} $. Once $ \bm{p^T} $ is optimized at the given position $ \bm{q} $, we optimize $ \bm{q} $ to maximize the sum capacity while considering the constraints in (\ref{eqn:3}). To this end, we first consider the problem of the sum capacity maximization regardless of the constraints in (\ref{eqn:3}). As the sum capacity is non-convex with respect to the FlyBS’s position, we provide a solution based on a local approximation of the sum capacity in the form of a radial function with respect to the FlyBS's position as elaborated in the following Lemma \ref{lem2}.

	\begin{lem}\label{lem2}
		The sum capacity $ C_{tot} $ is approximated as a radial function with respect to $\bm{q}[k]$ as:

		\begin{gather}
		C_{tot}[k]\approx W(\bm{p^T},k)-\zeta(\bm{p^T},k){\big|\big|\bm{q}[k]-\bm{S_0}(\bm{p^T},k)\big|\big|}^2,
		\label{eqn:15}
		\end{gather}
		\normalsize
		where the substitutions $W(\bm{p^T},k)$, $\zeta(\bm{p^T},k)$, and $\bm{S_0}(\bm{p^T},k)$ are constants with respect to $\bm{q}[k]$ as presented in the proof.		
	\end{lem}
	
	\textit{proof. Please see Appendix \ref{sec:apxB}}	
	
	

	

	According to (\ref{eqn:15}), the FlyBS achieves the maximum capacity at the location $ \bm{S_0} $. In addition, the sum capacity increases when the FlyBS’s distance to $ \bm{S_0} $ decreases. This helps to derive the FlyBS’s position for the constrained problem in (\ref{eqn:3}) in following way. The FlyBS’s position is updated to $ \bm{S_0} $ (as in (\ref{eqn:16})) if all constraints in (\ref{eqn:3}) are fulfilled, i.e., if $ \bm{S_0} $ lies inside the feasibility region denoted by $  \mathcal{R}_f  $. Otherwise, $ \bm{S_0} $ lies outside of $  \mathcal{R}_f  $ and the optimal position of the FlyBS (optimal with respect to (\ref{eqn:15})) is, then, the closest point from $  \mathcal{R}_f  $ to $ \bm{S_0} $. 
	If $ \bm{S_0} $ lies outside of $  \mathcal{R}_f  $, we refer to the derived spheres representing the constraints in (\ref{eqn:3}) (i.e., $ {sp}_j $  for $ j\in [1,N+2] $, see Section \ref{sec:3}) to find the closest point from  $ \mathcal{R}_f $ to $ \bm{S_0} $ and we provide a geometrical solution to determine the FlyBS's position as follows (also demonstrated in Algorithm \ref{alg:1}). 
	Due to the compactness of $ \mathcal{R}_f $,  the closest point of $ \mathcal{R}_f $ to $ \bm{S_0} $ lies on the boundary of $ \mathcal{R}_f $ belonging also to the border of at least one of the ($ N+2 $) spheres $ {sp}_j $. The closest point from any sphere $ {sp}_j $ to $ \bm{S_0} $ is determined by finding the intersection of $ {sp}_j $ and the straight line connecting $ \bm{S_0} $ to the center of $ {sp}_j $. Hence, we first find the closest point of each $ {sp}_j $  to $ \bm{S_0} $ (corresponding to line 1 in Algorithm \ref{alg:1}). 
	Next, we derive all mutual intersections (circles) of each pair of spheres $ {sp}_j $ and $ {sp}_k $ and we find the closest point from each of the intersection circles to $ \bm{S_0} $  (line 2 in Algorithm \ref{alg:1}- derivation steps not shown here to avoid cluttering, more details can be found in \cite{Eberly}). Similarly, we find the intersections of each sphere $ {sp}_j $ ($ j\in[1,N+2] $) with each of the planes $z=H_{min}$ and $z=H_{max}$ and then we find the closest points on those intersection circles to $ \bm{S_0} $ (lines 3 and 4 in Algorithm \ref{alg:1}).  
	After collecting all those closest points to $ \bm{S_0} $, we discard those collected points that do not fulfill all the conditions in (\ref{eqn:3}) (line 5 in Algorithm \ref{alg:1}). Last, in the remaining set of candidate points, the point with smallest distance to $ \bm{S_0} $ is the optimal position of the FlyBS (line 6 in Algorithm \ref{alg:1}).

	\begin{algorithm}[b]
		\textcolor{black}{\caption{ \textcolor{black}{\textcolor{Green}{}  Determination of the FlyBS positioning}}\label{alg:1} 
			Input: $sp_{j}$ ($j\in [1,N+2]$), and planes $z=H_{min}$, $z=H_{max}$\\ 
			$ \mathbf{\Lambda}=[]$: set of closest points to $ \bm{S_0} $ from the border of $ \mathcal{R}_f $ 
			\begin{algorithmic}[1]
				\State  $ \mathbf{\Lambda}\gets \mathbf{\Lambda}\cup \operatorname*{argmin}_{\bm{A}\in sp_{j}} ||\bm{S_0}-\bm{A}||$, $\forall j$
				\State  $ \mathbf{\Lambda}\gets \mathbf{\Lambda}\cup \operatorname*{argmin}_{\bm{D}\in sp_{j}\cap sp_k} ||\bm{S_0}-\bm{D}||$, $\forall j,k$
				\State  $ \mathbf{\Lambda}\gets \mathbf{\Lambda}\cup \operatorname*{argmin}_{\bm{B}\in sp_{j}\cap{z=H_{min}}} ||\bm{S_0}-\bm{B}||$, $\forall j$
				\State  $ \mathbf{\Lambda}\gets \mathbf{\Lambda}\cup \operatorname*{argmin}_{\bm{C}\in sp_{j}\cap{z=H_{max}}} ||\bm{S_0}-\bm{C}||$, $\forall j$
				\State  $ \mathbf{\Lambda}\gets \mathbf{\Lambda}-\{\bm{q}\in \mathbf{\Lambda }|\sim(\text{4b})\vee q\notin \cap_{j=1}^{N+2} sp_{j},  \} $
				\State $\bm{q} \gets \operatorname*{argmin}_{\bm{q}\in \mathbf{\Lambda}} ||\bm{S_0}-\bm{q}||$
			\end{algorithmic}
			\textbf{Output}: \textcolor{black}{FlyBS's position ($\bm{q}$)}}
	\end{algorithm}	\vspace{-0\baselineskip}
	\normalsize

	Once the FlyBS's position $ \bm{q} $ is updated, the power allocation $ \bm{p^T} $ is again optimized at the new $ \bm{q} $. The updated $ \bm{p^T} $ would change the spheres $ {sp}_j $ ($ j\in [1,N+2] $) and also $ \bm{S_0} $. Thus, the alternating optimization of  $ \bm{p^T} $ and $ \bm{q} $ continues until the FlyBS’s displacement at some iteration falls below a given threshold $ \epsilon $ or until the maximum number of iterations is reached. The complexity of finding the FlyBS’s position at each time step is $ O(N^4) $.  


	\section{Simulations and results}\label{sec:5}

	In this section, we present models and simulations adopted for a performance evaluation of the proposed solution, and we show gains of the proposal over state-of-the-art schemes.

	\subsection{Simulation scenario and models}\label{ssec:sim_scenario}

	We assume an area with a size of 600 x 600 m. Within this area, 60 to 180 nodes are dropped. A half of the nodes move based on a random-walk mobility model with a speed of 1 m/s. The other half of the nodes are randomly distributed into six clusters of crowds. The centers of three of the clusters move at a speed of 1 m/s, where each node in those clusters moves with a uniformly distributed speed of [0.6, 1.4] m/s with respect to the center of each cluster. The centers of the other three clusters move at a speed of 1.6 m/s with the speed of nodes uniformly distributed over [1.2, 2] m/s with respect to the center of cluster. 
	
	A total bandwidth of 100 MHz is selected \cite{Nikooroo2021}. Spectral density of noise is set to -174 dBm/Hz. The background interference is set to  -100 dBm. We set $ \alpha_i=2.4 $ for all nodes. The  allowed range for altitude of the FlyBS is $ [100,300] $ m, and the maximum transmission power limit $ P_{TX}^{max} $ is  1 W \cite{Alzenad2017}. A maximum speed of 25 m/s is assumed for the FlyBS. The maximum allowed propulsion power consumption (according to (4a)) is  set to $P_{pr,th}= 250$ W. Each simulation is of 1200 seconds duration. The results are averaged out over 100 simulation drops.

	In addition to our proposal, we show the performance of the following state-of-the-art solutions: $ i $) maximization of the minimum capacity of nodes (referred to as  \textit{MMC}) via the FlyBS’s positioning and the transmission power allocation, published in \cite{Valiulahi2020}, $ ii $) allocation of the transmission power to maximize an energy efficiency introduced in \cite{Muntaha2021} (referred to as  \textit{EEM}), $ iii $) allocation of the transmission power proposed in \cite{Muntaha2021} extended with K-means-based positioning of the FlyBS, as the solution in \cite{Muntaha2021} does not address the positioning; this approach is denoted as the extended \textit{EEM} (\textit{EEEM}).

	\begin{figure}[!t]
		\centering
		\includegraphics[width=2.4in]{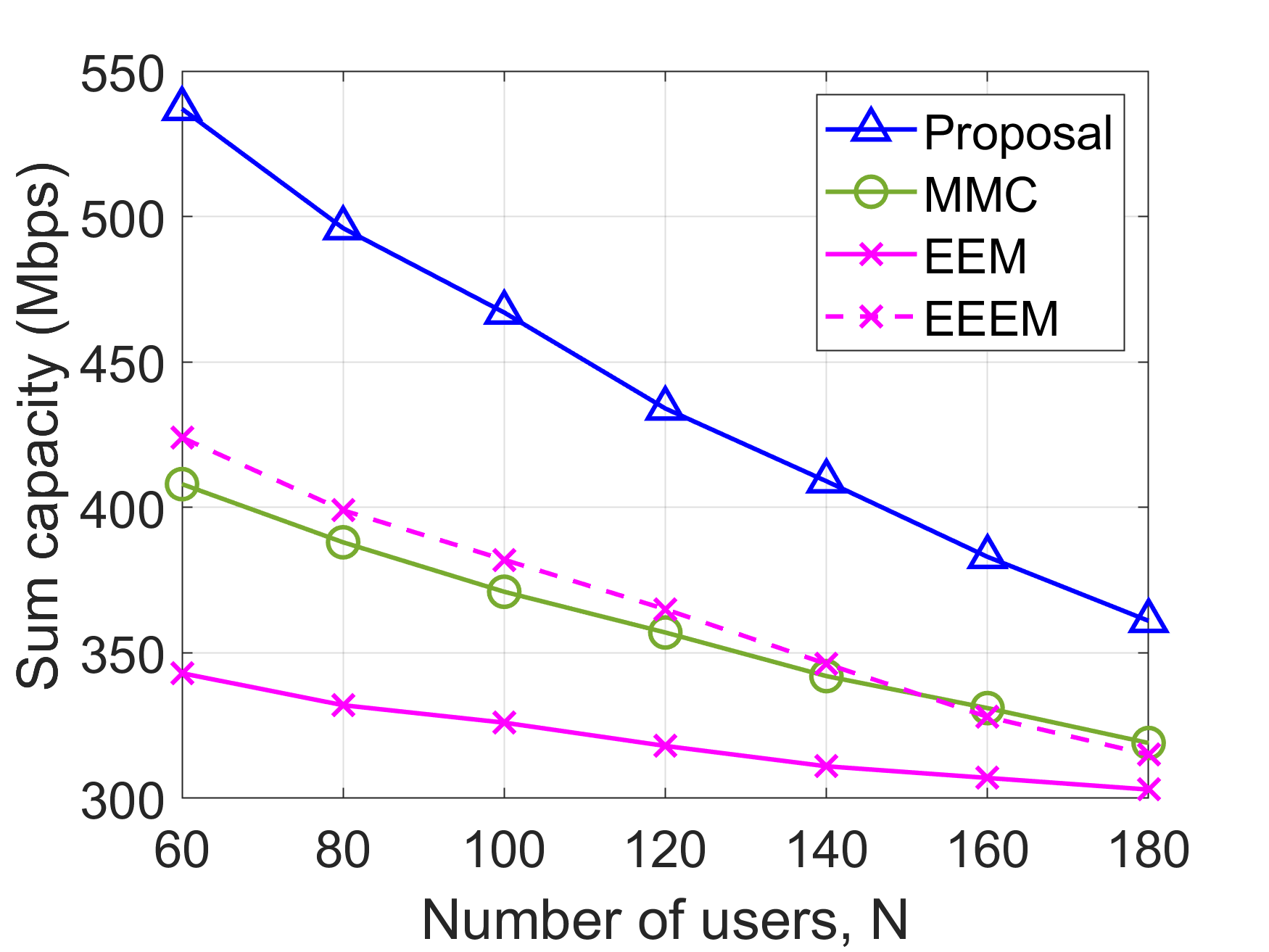}	
		\caption{Sum capacity vs. number of nodes for $C_{min}=$1 Mbps.} 
		\label{fig:sumC_N}
	\end{figure}\vspace{0\baselineskip}

	\subsection{Simulation results}\label{ssec:sim_result}

	In this subsection, we present and discuss simulation results. 
	Fig. \ref{fig:sumC_N} demonstrates the sum capacity versus number of nodes for $ C_i^{min}= 1$ Mbps for all nodes. The sum capacity decreases for larger numbers of nodes for all schemes, because the available bandwidth and the total transmission power is split among more nodes. However, our proposed solution enhances the sum capacity compared to state-of-the-art solutions \textit{MMC}, \textit{EEM}, and \textit{EEEM} by up to 26\%, 43\%, and 22\%, respectively.

	Figs. \ref{fig:sumC_Cmin_N=100} and \ref{fig:sumC_Cmin_N=180} show an impact of $C_{min}$ on the sum capacity for $ N= 100$ and $ N= 180$, respectively. 
	The maximum depicted $ C_i^{min} $ represents the largest $ C_i^{min} $ for which the feasible solution is found. Note that the value of $ C_i^{min} $ in \textit{MMC} is not set manually, but it is directly derived by the scheme itself. 
	For $ N= 100$ and $ N= 180$, the \textit{EEM} does not find a feasible solution for $ C_i^{min} $ larger than 2.2 Mbps and 0.8 Mbps, respectively, due to a lack of positioning of the FlyBS.    It is observed that the sum capacity decreases by $ C_i^{min} $ in the proposed solution, EEM, and EEEM. This is because the increasing $ C_i^{min} $ further limits the FlyBS’s allowed movement according to (\ref{eqn:5}) and, thus, the FlyBS can explore only a smaller feasibility region to optimize the sum capacity. 
	The proposed solution enhances the sum capacity with respect to \textit{MMC}, \textit{EEM}, and \textit{EEEM} by up to 25\%, 46\%, and 22\%, respectively, for $ N= 100$, and by up to 24\%, 26\%, and 15\%, respectively, for $ N= 180$.
	
	\begin{table}
		\begin{minipage}{0.44\linewidth}		
			
			\centering
			\includegraphics[width=1.75in]{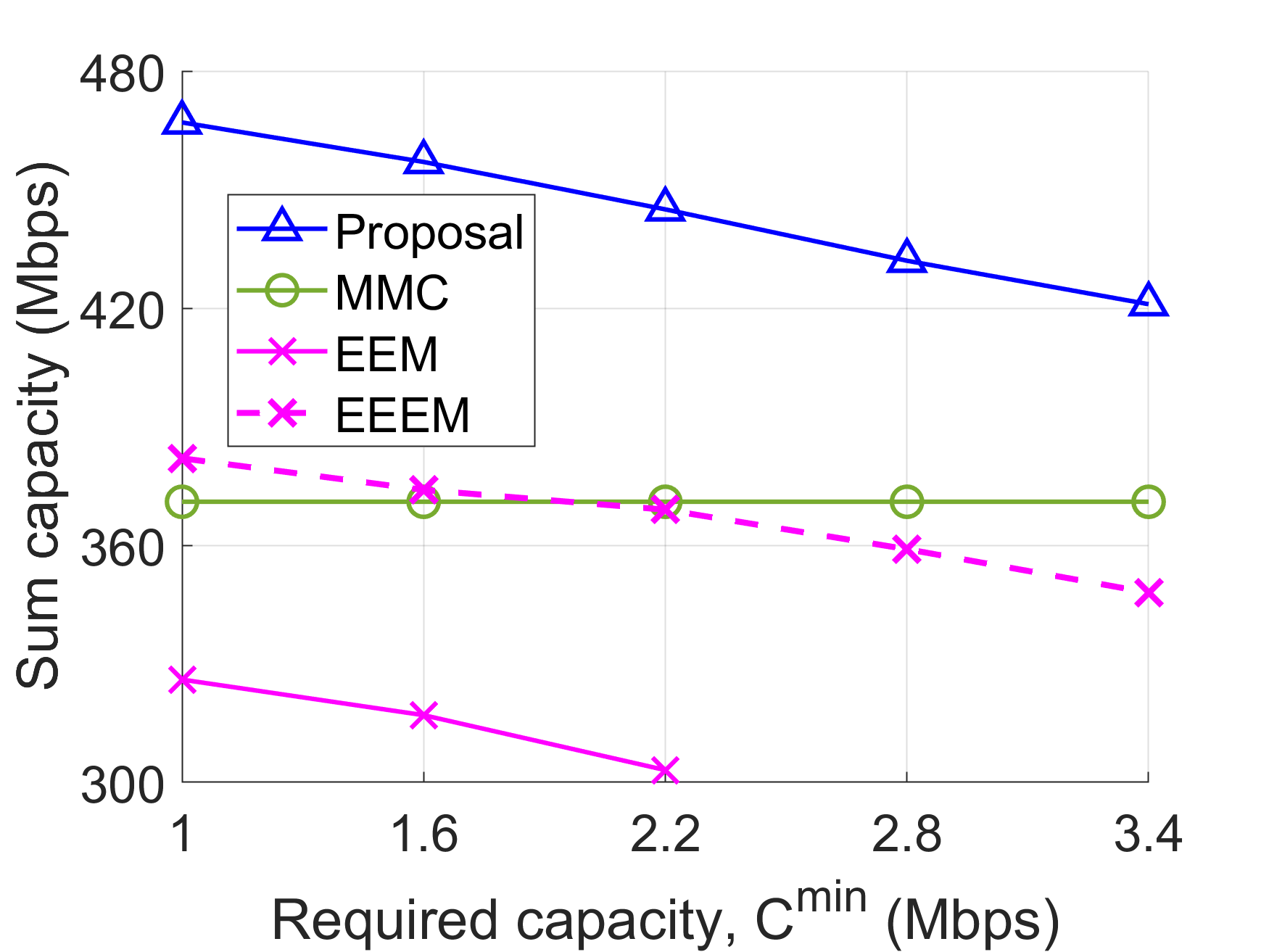}
			\captionof{figure}{\textcolor{black}{Sum capacity vs. $C_i^{min}$ for $N=100$.}}
			\label{fig:sumC_Cmin_N=100}
		\end{minipage}
		\hspace{0.35cm}
		\begin{minipage}{0.44\linewidth}
			\centering
			\includegraphics[width=1.75in]{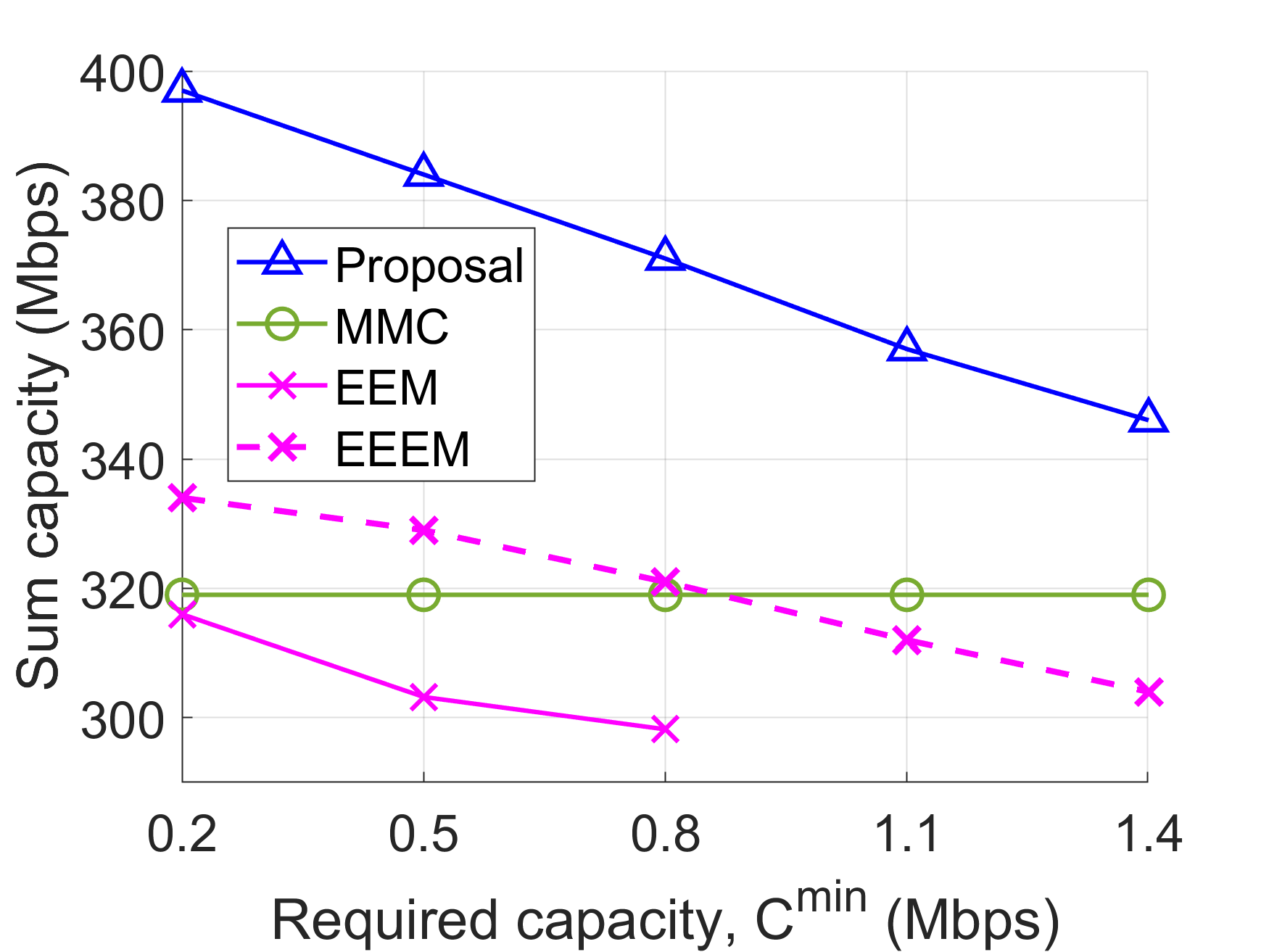}	
			\captionof{figure}{\textcolor{black}{Sum capacity vs. $C_i^{min}$  for $N=180$.}} 
			\label{fig:sumC_Cmin_N=180}	
		\end{minipage}
	\end{table}
	

	\begin{table}
	\begin{minipage}{0.46\linewidth}		
		
		\centering
		\includegraphics[width=1.75in]{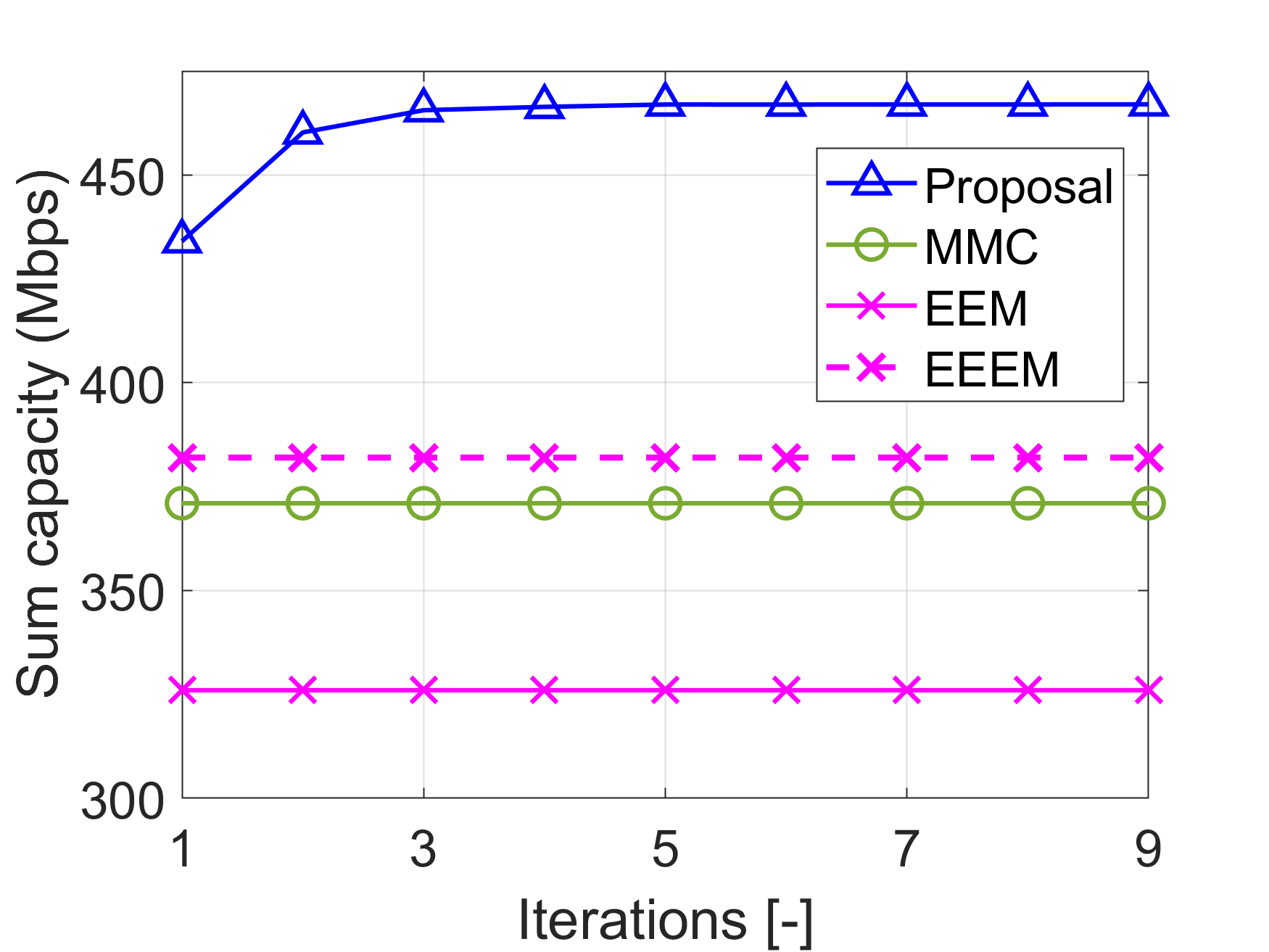}
		\captionof{figure}{Convergence of the proposed scheme for $N=100$.}
		\label{fig:sumC_iter_N=100}
	\end{minipage}
	\hspace{0.1cm}
	\begin{minipage}{0.46\linewidth}
		\centering
		\includegraphics[width=1.75in]{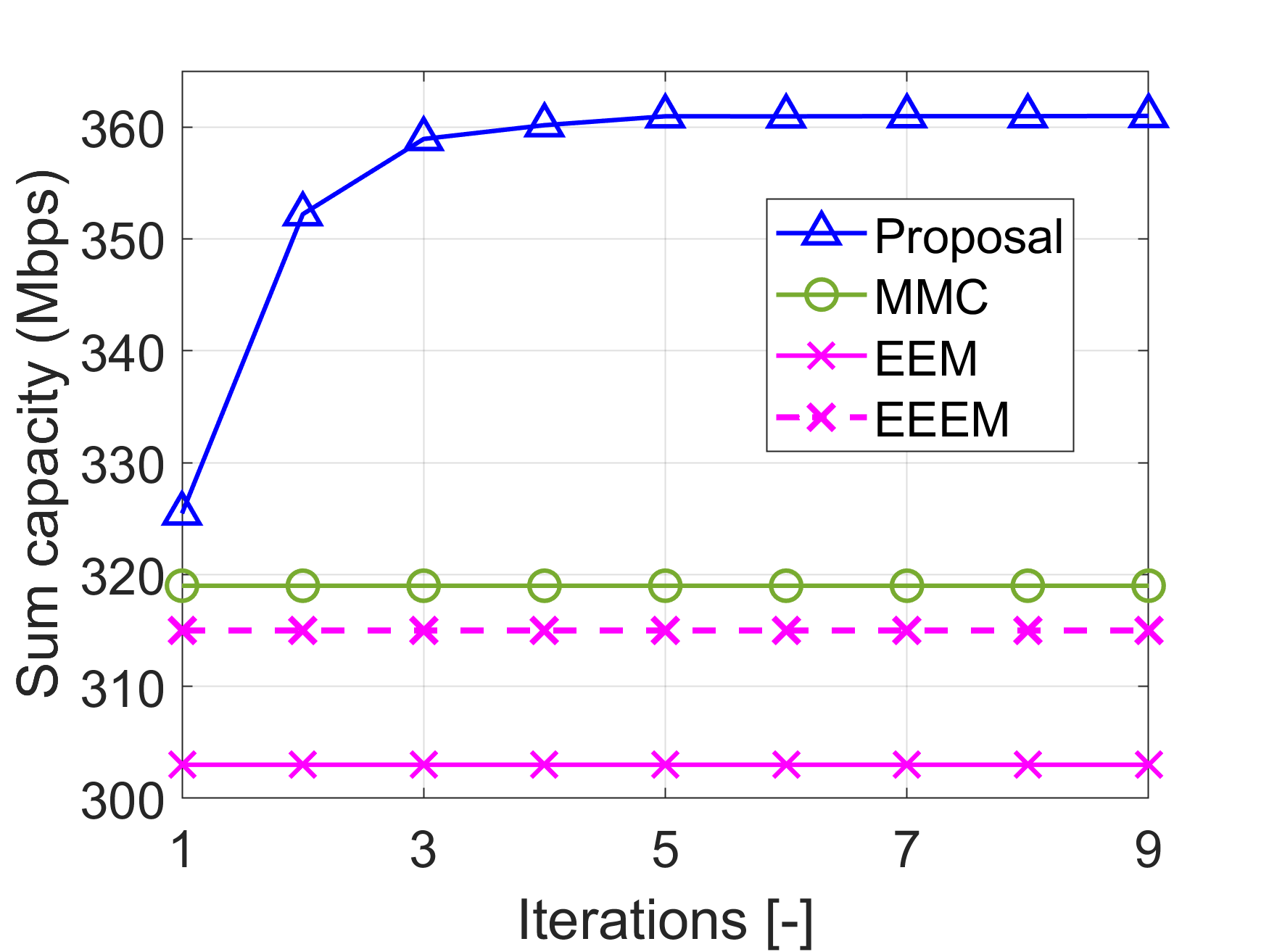}	
		\captionof{figure}{Convergence of the proposed scheme for $N=180$.} 
		\label{fig:sumC_iter_N=180}	
	\end{minipage}
\end{table}

	As our algorithm is iterative, we demonstrate its fast convergence in Figs. \ref{fig:sumC_iter_N=100} and \ref{fig:sumC_iter_N=180} by showing an evolution of the sum capacity over iterations of the FlyBS’s positioning and the transmission power allocation. The state-of-the-art schemes are not iterative, thus, their sum capacity is constant. Still, the proposed solution converges very fast, in about three iterations. Moreover, even the first iteration leads to a notably higher sum capacity comparing to all state-of-the-art solutions. This confirms that iterative approach does not limit feasibility and practical application of the proposed solution.
	

	\section{Conclusions}

	In this paper, we have provided a geometrical solution maximizing the sum capacity via a positioning of the FlyBS and an allocation of the transmission power to the nodes, while the minimum required capacity to each node is guaranteed. We have shown that the proposed solution enhances the sum capacity by tens of percent compared to state-of-the-art works. 
	
	In the future work, a scenario with multiple FlyBSs should be studied \textcolor{black}{along with related aspects, such as a management of interference among FlyBSs and an association of the nodes to the FlyBSs, should be addressed.}
	
	\begin{appendices}
		\section{Proof to Lemma \ref{lem1}} \label{sec:apxA} 
		\begin{proof}
			Using (\ref{eqn:4}), the constraint (4a) is rewritten as:
			\begin{gather}
			{{Q_i^{-1}d}_i}^{\alpha_i}[k](N_i+I)(2^\frac{C_i^{min}}{B_i}-1)\le p_i^T.
			\label{eqn:8}
			\end{gather}
			
			
			
			Then, the necessary condition to fulfill (4e) is that:
			\begin{gather}
			\sum_{i=1}^{N}{{{Q_i^{-1}d}_i}^{\alpha_i}(N_i+I)(2^\frac{C_i^{min}}{B_i}-1)}\le p_{max}^T.
			\label{eqn:10}
			\end{gather}

			\noindent	To derive an explicit form of (\ref{eqn:10}) in terms of the FlyBS’s position, we adopt the following inequality derived from linear Taylor approximation with respect to arbitrary $x$ and for $\eta\geq1$:	
			\begin{gather}
			{(a+x)}^\eta\geq{(a+\tau e\omega)}^\eta+\eta{(e+\tau e\omega)}^{\eta-1}(x-\tau a\omega),  
			\label{eqn:11}
			\end{gather}
			
			\noindent where $ \tau=\left\lfloor\frac{x}{a\sigma}\right\rfloor $, and $ \sigma $ is the approximation parameter such that choosing a smaller $ \sigma $ incurs a smaller  error. Hence, the approximation error and, thus, the gap to the optimal solution can be set arbitrarily close to zero by adopting a small enough  $ \sigma $.
			Using (\ref{eqn:11}), for the left-hand side in (\ref{eqn:10}) we write:
			\begin{gather}
			\sum_{i=1}^{N}{{Q_i^{-1}d}_i}^{\alpha_i}(N_i+I)(2^\frac{C_i^{min}}{B_i}-1)\ =\sum_{i=1}^{N}Q_i^{-1}(N_i+I)\times\nonumber\\(2^\frac{C_i^{min}}{B_i}-1)\times({(X-x_i)}^2+{(Y-y_i)}^2+{(H-z_i)}^2-
			H_{min}^2\nonumber\\+H_{min}^2)^\frac{\alpha_i}{2}\geq\sum_{i=1}^{N}{Q_i^{-1}(N_i+I)(2^\frac{C_i^{min}}{B_i}-1)}\times(\mu_i^{\frac{\alpha_i}{2}}-\nonumber\\\frac{\kappa_i\sigma\alpha_i}{2}\times\mu_i^{\frac{\alpha_i}{2}-1}H_{min}^2+(X^2+x_i^2-2Xx_i+Y^2+y_i^2-\nonumber\\2Yy_i+H^2+z_i^2
			-2Hz_i-H_{min}^2)\frac{\alpha_i}{2\mu_i^{-\frac{\alpha_i}{2}+1}})=\nonumber\\	(\sum_{i=1}^{N}\iota_i){\big|\big|\bm{q}[k]-\bm{\theta_0}(\bm{p^T},k)\big|\big|}^2+\chi,\label{eqn:12}
				\end{gather}
			\normalsize
			
			\noindent where
			
			\begin{gather} 
			\kappa_i=\left\lfloor\frac{\big|\big|{\bm{q}[k]-\bm{v_i}}\big|\big|^2-H_{min}^2}{H_{min}^2\sigma}\right\rfloor , \mu_i=H_{min}^2(1+\kappa_i\sigma), \nonumber\\ \iota_i={\frac{1}{2}Q_i^{-1}(N_i+I)(2^\frac{C_i^{min}}{B_i}-1)\alpha_i\mu_i^{\frac{\alpha_i}{2}-1}},
			\end{gather} 
			
			and
			\begin{gather} 
			\chi=-\sum_{i=1}^{N}{\iota_i(x_i^2+y_i^2+z_i^2)}+\nonumber\\\frac{{(\sum_{i=1}^{N}{\iota_ix_i})}^2+{(\sum_{i=1}^{N}{\iota_iy_i})}^2+{(\sum_{i=1}^{N}{\iota_iz_i})}^2}{\sum_{i=1}^{N}\iota_i}+\nonumber\\\sum_{i=1}^{N}{Q_i^{-1}(N_i+I)(2^\frac{C_i^{min}}{B_i}-1)}(\mu_i^{\frac{\alpha_i}{2}}-\frac{\kappa_i\sigma\alpha_i}{2}\mu_i^{\frac{\alpha_i}{2}-1}H_{min}^2).
			\label{eqn:13}
			\end{gather}
			\normalsize
			
		\noindent	Then, by incorporating (\ref{eqn:10}) and the right-hand side in (\ref{eqn:12}), Lemma 1 is proved.
		\end{proof}
		
		\section{Proof to Lemma \ref{lem2}} \label{sec:apxB} 
		\begin{proof}
			We use the following linear approximation (with respect to $ \mathrm{\Gamma} $) for arbitrary values of $ \mathrm{\Delta} $ and $ \mathrm{\Gamma} $:
			\begin{gather}
			\log_2(\mathrm{\Delta}+\mathrm{\Gamma})\approx\frac{1}{\text{ln}(2)}(\text{ln}(\mathrm{\Delta}+\mathrm{\Delta s\xi})+\frac{\mathrm{\Gamma}-s\Delta\xi}{\mathrm{\Delta}(1+s\xi)}),
			\label{eqn:17}
			\end{gather}
			
			\noindent where $ s=\left\lfloor\frac{\mathrm{\Gamma}}{\mathrm{\Delta\xi}}\right\rfloor $. Note that  the approximation error can be set arbitrarily close to zero by choosing small enough $ \xi $. 
			
			By taking $ \frac{p_i^T}{{{Q_id}_i}^{\alpha_i}(N_i+I)} $ as $ \mathrm{\Gamma} $ in (\ref{eqn:17}), the sum capacity is rewritten as:
			
			\small
			\begin{gather}
			C_{tot}[k]=\sum_{i=1}^{N}{B_i\log_2{\left(1+\frac{Q_ip_i^T}{{{d}_i}^{\alpha_i}(N_i+I)}\right)\approx}}\nonumber
			\end{gather}
			\normalsize
			\footnotesize
			\begin{gather}
			\sum_{i=1}^{N}\frac{B_ip_i^T{({(X-x_i)}^2+{(Y-y_i)}^2+(H-z_i)^2-H_{min}^2+H_{min}^2)}^\frac{-\alpha_i}{2}}{Q_i^{-1}(1+s_i\xi)(N_i+I)\text{ln}(2)}\nonumber
			\end{gather}
			\begin{gather}
			+\sum\nolimits_{i=1}^{N}\frac{B_i}{\text{ln}(2)}(\text{ln}(1+s_i\xi)-\frac{s_i\xi}{1+s_i\xi})\approx\nonumber\\
			\sum_{i=1}^{N}\frac{B_iQ_ip_i^T}{(1+s_i\xi)(N_i+I)\text{ln}(2)}\times(\mu_i^{\frac{\alpha_i}{2}}+\frac{\kappa_i\sigma\alpha_i}{2}\mu_i^{\frac{\alpha_i}{2}-1}H_{min}^2-\frac{\alpha_i\mu_i^{\frac{\alpha_i}{2}-1}}{2}\nonumber\\\times(X^2+x_i^2-2Xx_i+Y^2+y_i^2-2Yy_i+H^2+z_i^2-2Hz_i-H_{min}^2))\nonumber\\+\sum_{i=1}^{N}\frac{B_i}{\text{ln}(2)}(\text{ln}(1+s_i\xi)-\frac{s_i\xi}{1+s_i\xi})=W(\bm{p^T},k)-(\sum_{i=1}^{N}\varphi_i)\times((\nonumber\\{{X}-(\frac{\sum_{i=1}^{N}{\varphi_ix_i}}{\sum_{i=1}^{N}\varphi_i}))}^2+{({Y}-(\frac{\sum_{i=1}^{N}{\varphi_iy_i}}{\sum_{i=1}^{N}\varphi_i}))}^2+{({H}-(\frac{\sum_{i=1}^{N}{\varphi_iz_i}}{\sum_{i=1}^{N}\varphi_i}))}^2)\nonumber\\
			= W(\bm{p^T},k)-\zeta(\bm{p^T},k){\big|\big|\bm{q}[k]-\bm{S_0}(\bm{p^T},k)\big|\big|}^2,
			\label{eqn:18}\nonumber\\
			\end{gather}
			\normalsize
			
			\noindent where 
			
			\begin{gather}
			\bm{S_0}(\bm{p^T},k)=[\frac{\sum_{i=1}^{N}{\varphi_ix_i}}{\sum_{i=1}^{N}\varphi_i},\frac{\sum_{i=1}^{N}{\varphi_iy_i}}{\sum_{i=1}^{N}\varphi_i},\frac{\sum_{i=1}^{N}{\varphi_iz_i}}{\sum_{i=1}^{N}\varphi_i}],\label{eqn:16}
			\end{gather}
			and
			\begin{gather}
			W({\bm{p^T},t}_k)=\frac{{(\sum_{i=1}^{N}{\varphi_ix_i})}^2+{(\sum_{i=1}^{N}{\varphi_iy_i})}^2+{(\sum_{i=1}^{N}{\varphi_iz_i})}^2}{\sum_{i=1}^{N}\varphi_i}\nonumber\\-\sum_{i=1}^{N}{\varphi_i(x_i^2+y_i^2+z_i^2)}+\sum_{i=1}^{N}\frac{B_iQ_ip_i^T}{(1+s_i\xi)(N_i+I)\text{ln}(2)}\times \nonumber\\(\mu_i^{-\frac{\alpha_i}{2}}+\frac{\kappa_i\alpha_i\sigma H_{min}^2\mu_i^{-1-\frac{\alpha_i}{2}}}{2}+\frac{\alpha_i\mu_i^{\frac{\alpha_i}{2}-1}}{2}H_{min}^2)+\nonumber
			\end{gather}
			\begin{gather}
			\sum_{i=1}^{N}\frac{B_i}{\text{ln}(2)}(\text{ln}(1+s_i\xi)-\frac{s_i\xi}{1+s_i\xi}), \nonumber\\ s_i=\left\lfloor\frac{Q_ip_i^T}{\sigma(N_i+I)d_i^{\alpha_i}}\right\rfloor,\quad \varphi_i=\frac{Q_iB_ip_i^T\alpha_i\mu_i^{-1-\frac{\alpha_i}{2}}}{2\left(N_i+I\right)\left(1+s_i\xi\right)\text{ln}\left(2\right)},\nonumber
			\end{gather}

			\normalsize

			\begin{gather}
			\zeta({\bm{p^T},t}_k)=\sum_{i=1}^{N}\varphi_i. \label{auxiliary_1}
			\end{gather}
			\normalsize

			This proves Lemma 2.
		\end{proof}

	\end{appendices}
	
	

	


\begin{thebibliography}{00}
		
		\bibitem{Li2019} B. Li, et al, "UAV Communications for 5G and Beyond: Recent Advances and Future Trends," {\em IEEE Internet of Things Journal}, vol. 6, no. 2,  April 2019.
		
		\bibitem{Mach2022} P. Mach, et al, "Power Allocation, Channel Reuse, and Positioning of Flying Base Stations With Realistic Backhaul," {\em IEEE Internet of Things Journal}, vol. 9, no. 3, pp. 1790-1805, 1 Feb.1, 2022.
		
			\bibitem{Nikooroo2022TNSE} M. Nikooroo and Z. Becvar, "Optimization of Total Power Consumed by Flying Base Station Serving Mobile Users," {\em IEEE Trans. Netw. Sci. Eng.}, Early Access, 2022.
				\bibitem{Mozaffari2016} M. Mozaffari, et al, "Mobile Internet of Things: Can UAVs Provide an Energy-Efficient Mobile Architecture?," {\em IEEE  GLOBECOM}, 2016.
			
			\bibitem{Spyridis2021} Y. Spyridis, et al, "Towards 6G IoT: Tracing Mobile Sensor Nodes with Deep Learning Clustering in UAV Networks", {\em Sensors}, vol. 21, 2021.
			\bibitem{Esrafilian} O. Esrafilian, R. Gangula and D. Gesbert, "Learning to Communicate in UAV-Aided Wireless Networks: Map-Based Approaches," {\em IEEE Internet Things J.}, vol. 6, no. 2, 2019.
			
		\bibitem{Ahmed2020} S. Ahmed, et al, "Energy-Efficient UAV-to-User Scheduling to Maximize Throughput in Wireless Networks," {\em IEEE Access}, vol. 8, 2020.
		
		
		
		\bibitem{Ji2020} J. Ji, et al, "Joint Cache Placement, Flight Trajectory, and Transmission Power Optimization for Multi-UAV Assisted Wireless Networks," {\em IEEE Transactions on Wireless Communications}, vol. 19, no. 8, 2020.
		
		
		
		\bibitem{Wei2020} Z. Wei, et al, "Capacity of Unmanned Aerial Vehicle Assisted Data Collection in Wireless Sensor Networks," {\em IEEE Access}, vol. 8, 2020. 
		\bibitem{Hua2020_TCOM} M. Hua, et al, "3D UAV Trajectory and Communication Design for Simultaneous Uplink and Downlink Transmission," {\em IEEE Transactions on Communications}, vol. 68, no. 9, 2020. 
		\bibitem{Hua2020_WCL} M. Hua, et al, "Throughput Maximization for Full-Duplex UAV Aided Small Cell Wireless Systems," {\em IEEE Wireless Communications Letters}, vol. 9, no. 4, 2020.
		\bibitem{Li2021} B. Li, et al, "Full-Duplex UAV Relaying for Multiple User Pairs," {\em IEEE Internet of Things Journal}, vol. 8, no. 6, 2021.
		\bibitem{Xie2020}L. Xie, J. Xu and Y. Zeng, "Common Throughput Maximization for UAV-Enabled Interference Channel With Wireless Powered Communications," {\em IEEE Transactions on Communications}, vol. 68, no. 5, pp. 3197-3212, May 2020.
		
		\bibitem{Tun2021}Y. K. Tun, et al, "Energy-Efficient Resource Management in UAV-Assisted Mobile Edge Computing," {\em IEEE Communications Letters}, vol. 25, no. 1, pp. 249-253, Jan. 2021.
		
		\bibitem{Shi2020}L. Shi and S. Xu, "UAV Path Planning With QoS Constraint in Device-to-Device 5G Networks Using Particle Swarm Optimization," {\em IEEE Access}, vol. 8, pp. 137884-137896, 2020.
		\bibitem{Ishigami2020} M. Ishigami and T. Sugiyama, "A Novel Drone's Height Control Algorithm for Throughput Optimization in Disaster Resilient Network," {\em IEEE Transactions on Vehicular Technology}, vol. 69, no. 12, 2020. 
		\bibitem{Chen2020} R. Chen, et al, "Multi-UAV Coverage Scheme for Average Capacity Maximization," {\em IEEE Communications Letters}, vol. 24, no. 3, 2020.	
		
		\bibitem{Zhang2021} W. Zhang, et al, "Three-Dimension Trajectory Design for Multi-UAV Wireless Network With Deep Reinforcement Learning," {\em IEEE Transactions on Vehicular Technology}, vol. 70, no. 1, 2021.
		
		\bibitem{Muntaha2021} S. T. Muntaha, et al, "Energy Efficiency and Hover Time Optimization in UAV-Based HetNets," {\em IEEE Transactions on Intelligent Transportation Systems}, vol. 22, no. 8, 2021.
		
		\bibitem{Valiulahi2020} I. Valiulahi and C. Masouros, "Multi-UAV Deployment for Throughput Maximization in the Presence of Co-Channel Interference," in {\em IEEE Internet of Things Journal}, vol. 8, no. 5, 2020.
		
		
		
		
		
		
		
		
		
	
	
			\bibitem{Zeng2019} Y. Zeng, J. Xu, and R. Zhang, “Energy Minimization for Wireless Communication With Rotary-Wing UAV,” in {\em IEEE Trans. Wireless Commun.}, Vol. 18, No. 4, April 2019.
		
		\bibitem{Nikooroo2021} M. Nikooroo and Z. Becvar, "Optimal Positioning of Flying Base Stations and Transmission Power Allocation in NOMA Networks," {\em IEEE Trans. Wireless Commun.}, vol. 21, no. 2, pp. 1319-1334, Feb. 2022.
		
		
		
		
		
		
		
		
		
		
		
		
		
		\bibitem{Alzenad2017} M. Alzenad, et al, "3-D Placement of an Unmanned Aerial Vehicle Base Station (UAV-BS) for Energy-Efficient Maximal Coverage," {\em IEEE Wireless Communications Letters}, vol. 6, no. 4, pp. 434-437, Aug. 2017.
		
	
	
		\bibitem{Eberly} D. Eberly, "Distance to Circles in 3D
		", Geometric Tools, https://www.geometrictools.com/Documentation//DistanceToCircle3.pdf.
	
		
	\end{thebibliography}
\end{document}